\documentclass[twocolumn,english,prl,floatfix,superscriptaddress,showpacs]{revtex4}

\usepackage[T1]{fontenc}

\usepackage{amsmath,amsfonts,amssymb,color,epsfig,graphics,graphicx,latexsym,theorem,url,verbatim}

\usepackage{mathptmx}

\newcommand{\ket}[1]{|#1\rangle}

\let\theta\vartheta

\def\C{{\mathbb{C}}}

\newtheorem{theorem}{Theorem}
\newtheorem{proposition}[theorem]{Proposition}

\begin{document}

\title{Experimental Detection of Entanglement Polytopes via Local Filters}

\author{Yuanyuan Zhao}               %
\affiliation{Key Laboratory of Quantum Information, University of Science and Technology of China, CAS, Hefei, 230026, People's Republic of China}
\affiliation{Synergetic Innovation Center of Quantum Information and Quantum Physics, University of Science and Technology of China, Hefei, Anhui 230026, People's Republic of China}

\author{Markus Grassl}           %
\affiliation{Institut f\"ur Optik, Information und Photonik, Universit\"at Erlangen-N\"urnberg, Erlangen, Germany}
\affiliation{Max-Planck-Institut f\"ur die Physik des Lichts, Erlangen, Germany}

\author{Bei Zeng}               %
\affiliation{Department of Mathematics \&
Statistics, University of Guelph, Guelph, Ontario, Canada} %
\affiliation{Institute for Quantum Computing, University of Waterloo,
Waterloo, Ontario, Canada}

\author{Guoyong~Xiang}              %
\email{gyxiang@ustc.edu.cn}
\affiliation{Key Laboratory of Quantum Information, University of Science and Technology of China, CAS, Hefei, 230026, People's Republic of China}
\affiliation{Synergetic Innovation Center of Quantum Information and Quantum Physics, University of Science and Technology of China, Hefei, Anhui 230026, People's Republic of China}

\author{Chao Zhang}              %
\affiliation{Key Laboratory of Quantum Information, University of Science and Technology of China, CAS, Hefei, 230026, People's Republic of China}
\affiliation{Synergetic Innovation Center of Quantum Information and Quantum Physics, University of Science and Technology of China, Hefei, Anhui 230026, People's Republic of China}

\author{Chuanfeng Li}              %
\affiliation{Key Laboratory of Quantum Information, University of Science and Technology of China, CAS, Hefei, 230026, People's Republic of China}
\affiliation{Synergetic Innovation Center of Quantum Information and Quantum Physics, University of Science and Technology of China, Hefei, Anhui 230026, People's Republic of China}

\author{Guangcan Guo}              %
\affiliation{Key Laboratory of Quantum Information, University of Science and Technology of China, CAS, Hefei, 230026, People's Republic of China}
\affiliation{Synergetic Innovation Center of Quantum Information and Quantum Physics, University of Science and Technology of China, Hefei, Anhui 230026, People's Republic of China}
\
\begin{abstract}
%
Entanglement polytopes result in finitely many types of entanglement
that can be detected by only measuring single-particle spectra.  With
high probability, however, the local spectra lie in more than one
polytope, hence providing no information about the entanglement type.
To overcome this problem, we propose to additionally use local
filters.  We experimentally demonstrate the detection of entanglement
polytopes in a four-qubit system.  Using local filters we can
distinguish the entanglement type of states with the same single
particle spectra, but which belong to different polytopes.
\end{abstract}

\maketitle

Entanglement among quantum systems is a kind of quantum correlation
that is stronger than any possible classical
correlation~\cite{horodecki2009quantum}.  At the fundamental
  level, entanglement is the very mystery of quantum mechanics; and at
  the practical level,  it can be used as a physical resource to perform computation and communication tasks that are impossible for classical systems.
  Essentially, entanglement comes from the
tensor product structure of the Hilbert space of $N$ systems--qubits
in our case.  An $N$-qubit quantum state $\ket{\Psi_{N}}$ is entangled
if it can not be factored into products of quantum states of each of
the qubits.

One central question regarding entanglement is that how
$\ket{\Psi_{N}}$ may be entangled and how to detect that feature in
practice.  An obvious fact is that the parameters needed to specify
$\ket{\Psi_N}$ grows exponentially with $N$.  A natural idea to
eliminate some of the free parameters is to take that two states
$\ket{\Psi_N}$ and $\ket{\Phi_N}$ have similar entanglement features
if they can be connected by some single-qubit operations, for
instance, local unitary (LU) transformation
\cite{acin2001three,CDGZ14,CCDZ15} or stochastic local operation
combined with classical communication (SLOCC)
\cite{dur2000three,verstraete2002four,OD07}.

For $N=2$, the Schmidt decomposition tells us that
$\ket{\Psi_2}=\lambda_0\ket{00}+\lambda_1\ket{11}$ up to LU, with
$\lambda_0\ge\lambda_1$ and $\lambda_0^2+\lambda_1^2=1$. Different
$\lambda_0\in[1/\sqrt{2},1]$ corresponds to different LU classes of
entanglement, which are in fact infinitely many.  Up to SLOCC,
however, there is only one class of entangled states which contains
the EPR pair with $\lambda_0=1/\sqrt{2}$.  For $N=3$, up to SLOCC,
there are only two types of entanglement: the GHZ-type state and the
$W$-type state~\cite{dur2000three}. These two types can be
distinguished by a quantity called $3$-tangle, which is however not a
single-copy observable and hence cannot be directly measured in
experiment~\cite{coffman2000distributed} (that is, to get the
value of $3$-tangle, one either
needs to measure jointly on multiple copies of the states,
or needs a state tomography~\cite{guhne2009entanglement}).

For any $N>3$, SLOCC no longer results in a finite number of
entanglement types~\cite{verstraete2002four,OD07}.  Despite the efforts of
studying SLOCC classification of entanglement for $N>3$ systems, the
exponential growth of parameters with $N$ for describing
$\ket{\Psi_{N}}$ makes it hopeless to extract clear physical meanings
of these classifications. It is highly desired to coarse-grain
these classes such that we can grasp the key features of each
entanglement type. The concept of entanglement polytopes provides an
elegant idea to meet this need~\cite{walter2013entanglement}, where
for each $N$ there exists only finite number of types. More
importantly, the polytopes are directly detectable in experiments via
measuring only single-particle spectra of each
qubit \cite{walter2013entanglement,sawicki2012critical,sawicki2014convexity}.

In this work, we experimentally demonstrate the detection of
entanglement polytopes in a four-qubit system.  Unfortunately, it
turns out that different entanglement polytopes form a nested
hierarchy
\cite{walter2013entanglement,sawicki2012critical,sawicki2014convexity},
and they may have a large overlap.  If the vector of local spectra of
a state $\ket{\Psi_N}$ lies in an overlapping region, then we cannot
uniquely identify the polytope that $\ket{\Psi_N}$ belongs to (see
e.g. a recent experiment in which the states are chosen to be in
non-overlapping regions \cite{aguilar2014experimental}).

It turns out that for a randomly chosen three-qubit pure state
$\ket{\Psi_3}$, the probability that the vector of local spectra of
$\ket{\Psi_3}$ lies in the overlapping region of the $W$ and GHZ
polytopes is $\approx 94\%$.  In general, for a randomly chosen state
$N$-qubit state $\ket{\Psi_N}$, with high probability the vector of
local spectra falls in some overlapping region of polytopes (we
include a more detailed discussion of these probabilities in the
Appendix).  To overcome this difficulty, we use local
filters (see,
e.g. \cite{verstraete2001local,wang2006experimental,bai2008multipartite,campbell2009characterizing,bastin2009operational})
to effectively distinguish states with the same single particle
spectra, but which belong to different polytopes.

\textit{Entanglement Polytopes --} An $N$-qubit quantum state
$\ket{\Psi_N}$ is said to be convertible to another $N$-qubit state
$\ket{\Phi_N}$ via SLOCC if there exists a sequence of local
operations and classical communication that converts the state
$\ket{\Psi_N}$ to $\ket{\Phi_N}$ with nonzero probability. The states
$\ket{\Psi_N}$ and $\ket{\Phi_N}$ are said to be SLOCC equivalent if
$\ket{\Phi_N}$ is convertible to $\ket{\Phi_N}$ and vice versa.  It
has been shown that $\ket{\Psi_N}$ is SLOCC equivalent to
$\ket{\Phi_N}$ if and only if there exist invertible matrices $M_i$ for
$i=1,2,\ldots,N$ such that
\begin{equation}\label{eq:SLOCC}
\ket{\Psi_N}=M_1\otimes M_2\otimes\ldots\otimes M_N\ket{\Phi_N}.
\end{equation}
This SLOCC equivalence relation partitions all $N$-qubit pure states
into SLOCC equivalent classes, called the SLOCC orbits.

For an $N$-qubit state $\ket{\Psi_N}$, each of the single-particle
reduced density matrices $\rho_i$ for $i=1,\ldots,N$ has two
eigenvalues $\lambda_i^{\alpha},\lambda_i^{\beta}$ that are
normalized, i.e. $\lambda_i^{\alpha}+\lambda_i^{\beta}=1$. It suffices
to consider the maximum eigenvalue of $\rho_i$,
i.e. $\lambda_i^{\text{max}}=\max(\lambda_i^{\alpha},\lambda_i^{\beta})$,
with $\frac{1}{2}\leq\lambda_i^{\text{max}}\leq 1$. The
$N$-dimensional vector
$\vec{\lambda}=(\lambda_1^{\text{max}},\lambda_2^{\text{max}},\ldots,\lambda_N^{\text{max}})$
then corresponds to a point in $\mathbb{R}^N$.

It has been shown that for the closure of an $N$-qubit SLOCC orbit
$\bar{\mathcal{O}}$ of $\Psi_N$ (i.e. some $M_i$ in (\ref{eq:SLOCC})
possibly non-invertible), all the points $\vec{\lambda}$ of all
$\bigotimes_{i=1}^N M_i\ket{\Psi_N}\in\bar{\mathcal{O}}$ form a
polytope in $\mathbb{R}^N$, called the entanglement polytope of
$\bar{\mathcal{O}}$. Moreover, for any finite $N$, there are only
finitely many polytopes. This provides a natural classification of
entanglement for $N$-qubit states, which `coarse-grains' the
infinitely many SLOCC orbits (for $N>3$).  Since the entanglement
polytope for an SLOCC orbit is fully determined by the local spectra
of the states in the orbit, this offers an appealing experimental
approach for identifying the entanglement type for an $N$-qubit
system, for which only measurements on single particles are required.

As a simple example, for $N=3$, with only two kinds of SLOCC orbits
(GHZ and $W$-type states), there are two polytopes in
$\mathbb{R}^3$. The polytope $\mathcal{P}^{\text{GHZ}}$ corresponds to
the GHZ-type states, with vertices
$(1/2,1/2,1/2)$, $(1/2,1/2,1)$, $(1/2,1,1/2)$, $(1,1/2,1/2)$, $(1,1,1)$;
and $\mathcal{P}^{W}$ corresponds to the $W$-type states, with vertices
$(1,1,1)$, $(1/2,1/2,1)$, $(1/2,1,1/2)$, $(1,1/2,1/2)$.
Obviously, $\mathcal{P}^{W}\subset\mathcal{P}^{\text{GHZ}}$, which
shows that entanglement polytopes for different SLOCC orbits may
overlap.

In practice, for a state $\ket{\Psi_N}$, while a point
$\vec{\lambda}$ may clearly distinguish its entanglement type, if
$\vec{\lambda}$ is in an overlapping region of two polytopes, we fail
to get information on which entanglement type the state belongs to. In
the $N=3$ case, for instance, this means that a point
$\vec{\lambda}\in\mathcal{P}^{W}$ fails to distinguish the $W$-type
entanglement from the GHZ-type entanglement.  Unfortunately, for a
randomly chosen pure state of three qubits, with $\approx 94\%$
probability the corresponding $\vec{\lambda}$ falls into
$\mathcal{P}^{W}$.

Luckily, one can apply local filter operations $\bigotimes_{i=1}^N
M_i$ to the system to `move around' $\vec{\lambda}$, with the hope
that $\vec{\lambda}$ ends up in a non-overlapping area of
polytopes.  As demonstrated in the three-qubit case, this step becomes
crucial in practice when using the polytope method for detecting
entanglement types, as the probability of overlapping is high.

\textit{Four-qubit Polytopes --} Our experiments demonstrate the
detection of entanglement types of four-qubit states. In this case,
there are infinitely many SLOCC orbits. The full polytope, containing
$\vec{\lambda}$ for any four-qubit state, denoted by
$\mathcal{P}^{\text{full}}$, is spanned by the vertices
\begin{alignat*}{5}
&(\frac{1}{2},\frac{1}{2},\frac{1}{2},\frac{1}{2}),(1,1,1,1),\\
&(1,1,\frac{1}{2},\frac{1}{2}),(1,\frac{1}{2},1,\frac{1}{2}),(1,\frac{1}{2},\frac{1}{2},1),\\
&(\frac{1}{2},1,1,\frac{1}{2}),(\frac{1}{2},1,\frac{1}{2},1),(\frac{1}{2},\frac{1}{2},1,1),\\
&(1,\frac{1}{2},\frac{1}{2},\frac{1}{2}),(\frac{1}{2},1,\frac{1}{2},\frac{1}{2}),(\frac{1}{2},\frac{1}{2},1,\frac{1}{2}),(\frac{1}{2},\frac{1}{2},\frac{1}{2},1).
\end{alignat*}

\begin{figure}[hbpt]
\centerline{
\begin{picture}(220,85)
\put(20,50){\makebox(0,0){}}
\put(0,0){\framebox(40,80){$\ket{\Psi^{(i)}}$}}
\multiput(40,10)(0,20){4}{\line(1,0){19}}
\put(59,5){\framebox(2,10){}}
\put(59,17){\makebox(0,0)[b]{\footnotesize }}
\put(67,8){\makebox(0,0)[t]{$\vartheta_1$}}
\put(59,65){\framebox(2,10){}}
\put(59,77){\makebox(0,0)[b]{\footnotesize }}
\put(67,68,63){\makebox(0,0)[t]{$\vartheta_2$}}
\multiput(61,10)(0,20){4}{\line(1,0){18}}
\multiput(40,30)(0,20){2}{\line(1,0){60}}
\put(79,5){\framebox(2,10){}} \put(79,17){\makebox(0,0)[b]{}}
\put(85,8){\makebox(0,0)[t]{$\gamma$}}
\multiput(81,10)(0,20){4}{\line(1,0){19}}
\multiput(61,30)(0,20){3}{\line(1,0){39}}
\multiput(100,10)(0,20){4}{\oval(10,10)[r]}
\multiput(100,5)(0,20){4}{\line(0,1){10}}
\multiput(105,10)(0,20){4}{\vector(1,0){10}}
\put(117,70){\makebox(0,0)[l]{$\lambda^{\text{max}}_1=1$ (post-selection)}}
\put(117,50){\makebox(0,0)[l]{$\lambda^{\text{max}}_2$}}
\put(117,30){\makebox(0,0)[l]{$\lambda^{\text{max}}_3$}}
\put(117,10){\makebox(0,0)[l]{$\lambda^{\text{max}}_4$}}
\end{picture}}
\caption{Circuit diagram of the experimental setup.}
\label{fig:setup}
\end{figure}
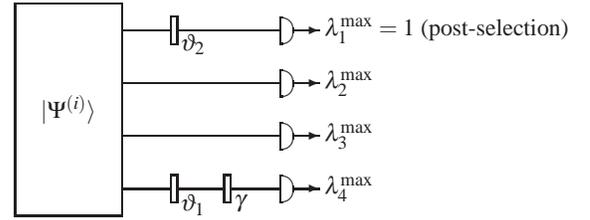

Up to permutation of the qubits, there are $6$ other polytopes inside
$\mathcal{P}^{\text{full}}$, which may also be mutually overlapping.
Similar as in the $N=3$ case, for a randomly chosen four-qubit state,
the chance that $\vec{\lambda}$ lies in an overlapping region is high
(for details, see the Appendix).  Therefore, we have to
apply local operations to `move around' $\vec{\lambda}$. The proposed
experiment is given by the diagram in Fig.~\ref{fig:setup}.

Here ${\vartheta}_i$ for $i=1,2$ denotes a unitary local
transformation ${\rm U}_{\vartheta_i}$ of the form
\begin{equation}
{\rm U}_{\vartheta_i}=
\begin{pmatrix}
\cos{\vartheta_i} & -\sin{\vartheta_i}\\ \sin{\vartheta_i} &
\cos{\vartheta_i}
\end{pmatrix}
\begin{pmatrix}
1 & 0\\ 0 & -1
\end{pmatrix}
\begin{pmatrix}
\cos{\vartheta_i} & \sin{\vartheta_i}\\ -\sin{\vartheta_i} &
\cos{\vartheta_i}
\end{pmatrix},\nonumber
\end{equation}
and ${\gamma}$ denotes a non-unitary local transformation
\begin{equation}
{\rm A}_{\gamma}=
\begin{pmatrix}
1 & 0 \\ 0 &\gamma
\end{pmatrix}, \nonumber
\end{equation}

\begin{figure*}[hbpt]
\begin{center}
\includegraphics [width=12cm,height=8cm]{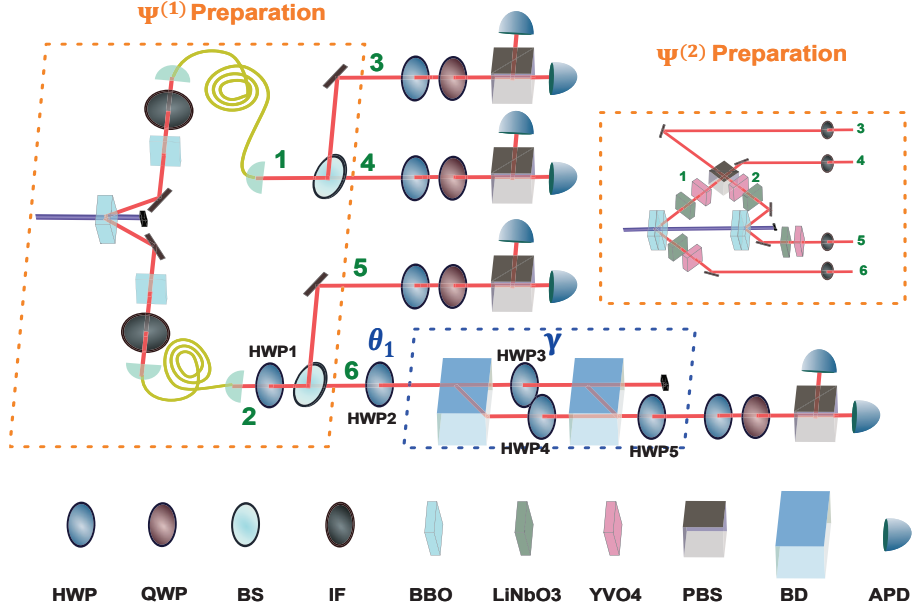}
\end{center}
\caption{ Detailed configurations for preparing the states
  $\Psi^{(1)}$ and $\Psi^{(2)}$ and for realizing the operator
  $A_{\gamma}$ are shown in the orange boxes and the blue box, where
  HWP3, HWP4 and HWP5 are rotated by $\frac{1}{2}(\arcsin{\gamma})$,
  $\frac{\pi}{4}$, and $\frac{\pi}{4}$ respectively.  The unitary
  operator $U_{\theta_{1}}$ is realized by HWP2 at specific angles. A
  QWP and a HWP in front of a polarization beam displacer (PBS) in
  each mode are used to implement the measurement in different bases
  for the standard state tomography.  Post-selection in some basis of
  qubit $1$ (the state of the photon in mode `$3$') is realized by
  collecting only photons in one of the output modes of the PBS.  The
  indices in the figure denote the spatial modes. }
 \label{fig:exp_setup}
\end{figure*}

In Fig.~\ref{fig:setup}, two of the qubits encounter non-unitary local
transformations: qubit $1$ is measured in some basis and
post-selected, resulting in $\lambda^{\text{max}}_1=1$; qubit $4$ is
going through a filter operation given by ${\rm A}_{\gamma}$. In the
most general case, one can also apply local filter operations (or
measurements) on the other qubits.  However, a single filter (or
measurement) may already suffice to `move around' $\vec{\lambda}$ to
non-overlapping regions of the polytopes, depending on the input state
$\ket{\Psi^{(i)}}$.

\textit{Experimental setup --} In our experiments, two different four-qubit states $\ket{\Psi^{(1)}}$ and
$\ket{\Psi^{(2)}}$ are prepared, where
\begin{eqnarray}
\ket{\Psi^{(1)}}&=&\frac{\sqrt{3}}{3}(\ket{HHHH}+\ket{VVVV})\nonumber\\ &+&\frac{\sqrt{3}}{6}(\ket{HV}+\ket{VH})(\ket{HV}+\ket{VH}),
\end{eqnarray}
and $\ket{\Psi^{(2)}}$ is the four-qubit GHZ state
\begin{equation}
\ket{\Psi^{(2)}}=\frac{\sqrt{2}}{2}(\ket{HHHH}+\ket{VVVV}).
\end{equation}
The qubits are encoded by horizontal $\ket{H}$ and vertical $\ket{V}$
polarization.  The goal is to determine the entanglement type for each
of the input state using the polytope method.  For both
$\ket{\Psi^{(1)}}$ and $\ket{\Psi^{(2)}}$, we have
$\vec{\lambda}=(\frac{1}{2},\frac{1}{2},\frac{1}{2},\frac{1}{2})$. That
is, local spectra do not tell them apart, hence local filter
operations are needed to `move around' $\vec{\lambda}$.

The SLOCC orbit of the four-qubit GHZ state $\ket{\Psi^{(2)}}$
correspond to the full polytope $\mathcal{P}^{\text{full}}$. However,
the state $\ket{\Psi^{(1)}}$ corresponds to a smaller polytope
$\mathcal{P}^{s}\subset\mathcal{P}^{\text{full}}$ with vertices
\begin{alignat*}{5}
&(\frac{1}{2},\frac{1}{2},\frac{1}{2},\frac{1}{2}),(1,1,1,1),\\
&(1,1,\frac{1}{2},\frac{1}{2}),(1,\frac{1}{2},1,\frac{1}{2}),(1,\frac{1}{2},\frac{1}{2},1),\\
&(\frac{1}{2},1,1,\frac{1}{2}),(\frac{1}{2},1,\frac{1}{2},1),(\frac{1}{2},\frac{1}{2},1,1).
\end{alignat*}
The smaller polytope $\mathcal{P}^{s}$ is characterized by the
additional constraint
\begin{equation}\label{eq:condition}
f(\vec{\lambda})=
-\lambda_1^{\text{max}}+\lambda_2^{\text{max}}+\lambda_3^{\text{max}}+\lambda_4^{\text{max}}
\ge 1
\end{equation}
and all permutations of it.

Our experimental setup for the states $\ket{\Psi^{(1)}}$ and
$\ket{\Psi^{(2)}}$ is shown in Fig.~\ref{fig:exp_setup}.  A
$390~\text{nm}$ femto-second pump light, frequency-doubled from a
$780~\text{nm}$ mode-locked Ti:sapphire pulsed laser (with the pulse
width about $150~\text{fs}$ and repetition rate $76~\text{MHz}$) was
used to pump the respective down-converter.  For the preparation of
$\ket{\Psi^{(1)}}$, a $2~\text{mm}$ type-II phase-matched BBO crystal
is used as down-converter to produce two pairs of entangled photons
\cite{kwiat1995new}, and two $1~\text{mm}$ BBO crystals are used to
compensate the birefringence of $o$-light and $e$-light in the
$2~\text{mm}$ BBO.  HWP1 rotates the polarization of the photons in
path `$2$' (horizontal to vertical and vertical to horizontal). Then
after the beam splitters (BS), the above two pairs of entangled
photons are transformed into the state $\ket{\Psi^{(1)}}$. In mode
`$6$', we use two beam displacers and three half wave plates (HWPs,
HWP3 is used for balancing the optical length of the two beams between
partdisplacers) to construct the local filter ${\rm A}_{\gamma}$.  For the
four-qubit GHZ state $\ket{\Psi^{(2)}}$ shown in the right part of
Fig. \ref{fig:exp_setup}, a cascaded sandwich beam-like BBO
entangled source \cite{Huang14} is used. A PBS combines the photons
from mode `$1$' and `$2$'.  We will get the four-qubit GHZ state
$\ket{\Psi^{(2)}}$ if there is one photon in each of the modes `$3$',
`$4$', `$5$', and `$6$' \cite{pan2000experimental}.

\textit{Results --} We first perform full quantum state tomography to
reconstruct the density matrix of $\ket{\Psi^{(1)}}$ and
$\ket{\Psi^{(2)}}$, the fidelity of which are $0.9422\pm0.0036$ and
$0.9001\pm0.0038$. Then we collect data from each mode to obtain the
corresponding single-qubit density matrix and calculate their local
spectra for both states. As shown in Table \ref{tab:4qubit}, the local
spectra of $\ket{\Psi^{(1)}}$ and $\ket{\Psi^{(2)}}$ are almost
identical, so we can not distinguish their entanglement polytopes.

\begin{table}[hbpt]
\begin{align*}
\begin{array}{|c|c|c|c|c|c|c|}
    \hline
    \text{state} & \lambda_{1}^{\text{max}} & \lambda_{2}^{\text{max}} & \lambda_{3}^{\text{max}} & \lambda_{4}^{\text{max}} &f(\vec{\lambda})\\
    \hline
    \ket{\Psi^{(1)}} & 0.529(4) & 0.514(4) &  0.540(4) &  0.530(4) &  1.056(8)\\
    \hline
    \ket{\Psi^{(2)}} & 0.521(4) & 0.524(4) &  0.535(4) &  0.525(4) &  1.062(8)\\
    \hline
\end{array}
\end{align*}
\caption{The local spectra $\lambda_{1}^{\text{max}}$,
  $\lambda_{2}^{\text{max}}$, $\lambda_{3}^{\text{max}}$,
  $\lambda_{4}^{\text{max}}$ together with $f(\vec{\lambda})$ for the states
  $\Psi^{(1)}$ and $\Psi^{(2)}$.  The uncertainties inside the
  brackets are obtained by Monte Carlo simulation ($1000$ steps).}
\label{tab:4qubit}
\end{table}

\begin{table}[hbt!]
\begin{align*}
\begin{array}{|c|c|c||c|c|c||c|}
    \hline
        &  \theta_{1} & \gamma & \lambda_{2}^{\text{max}} & \lambda_{3}^{\text{max}} & \lambda_{4}^{\text{max}} &f(\vec{\lambda}) \\
    \hline
    a   & \pi/8   & 1/\sqrt{2}  &  0.609(10) &  0.831(9) &  0.701(10) &  1.141(18)\\
    \hline
    b   & \pi/8   & 1/\sqrt{3}  & 0.557(10) &  0.850(9) &  0.614(10) &  1.021(17)\\
    \hline
    c   & \pi/8   & 1/\sqrt{5}  & 0.603(10) &  0.875(9) &  0.553(10) &  1.032(18)\\
    \hline
    d   & 3\pi/32 & 1/\sqrt{5}  & 0.713(9) &  0.883(9) &  0.717(8) &  1.313(17)\\
    \hline
    e   & 0       & 1           & 0.657(9) &  0.848(8) &  0.857(8) &  1.362(17)\\
    \hline
    f   & 0       & 1           & 0.525(9) &  0.544(8) &  0.516(8) &  0.584(19)\\
    \hline
\end{array}
\end{align*}
\caption{Setting of the parameters $\theta_{1}$ and $\gamma$ for the
  data points labeled `$a \thicksim f$', together with the measured
  local spectra $\lambda_{2}^{\text{max}}$,
  $\lambda_{3}^{\text{max}}$, $\lambda_{4}^{\text{max}}$ and the
  resulting value of $f(\vec{\lambda})$. The uncertainties inside the
  brackets are obtained by Monte Carlo simulation ($1000$ steps). }
\label{tab:spectra}
\end{table}

To distinguish the entanglement polytopes of $\ket{\Psi^{(1)}}$ and
$\ket{\Psi^{(2)}}$, we then try to move $\vec{\lambda}$ out of the
smaller polytope $\mathcal{P}^{s}$ using local filters, as illustrated
in Fig.~\ref{fig:setup}.  We fix $\theta_2=-\pi/8$, and then measure
the first qubit in the computational basis.  By post-selection we have
$\lambda_1^{\text{max}}=1$.  For each setting of ${\vartheta}_1$ and
$\gamma$, we perform tomography of the qubits $2$, $3$, and $4$ to
determine the values of $\lambda_2^\text{max}$,
$\lambda_3^\text{max}$, and $\lambda_4^\text{max}$ (see Table
\ref{tab:spectra}).  The smaller polytope $\mathcal{P}^s$ is
characterized by $f(\vec{\lambda})\ge 1$.

\begin{figure}[hbpt]
\begin{center}
\includegraphics [width= 0.65\columnwidth]{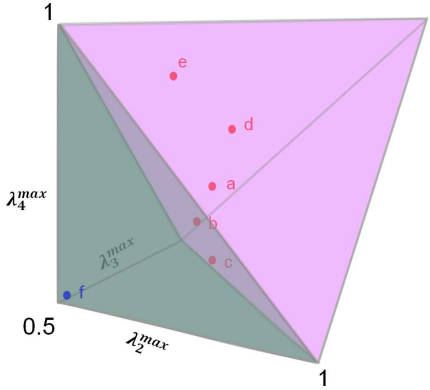}
\end{center}
\caption{Three-dimensional polytopes.
The pink region and the blue region represent the polytope of $\ket{\Psi^{(1)}}$
and $\ket{\Psi^{(2)}}$ respectively. Experimental data `$a \thicksim e$' is for $\ket{\Psi^{(1)}}$
while `$f$' is for $\ket{\Psi^{(2)}}$. Error bars are too small to identify (see Table~\ref{tab:spectra}). }
\label{fig:resultp}
\end{figure}

\begin{figure}[hbpt]
\begin{center}
\includegraphics [width= 0.88\columnwidth]{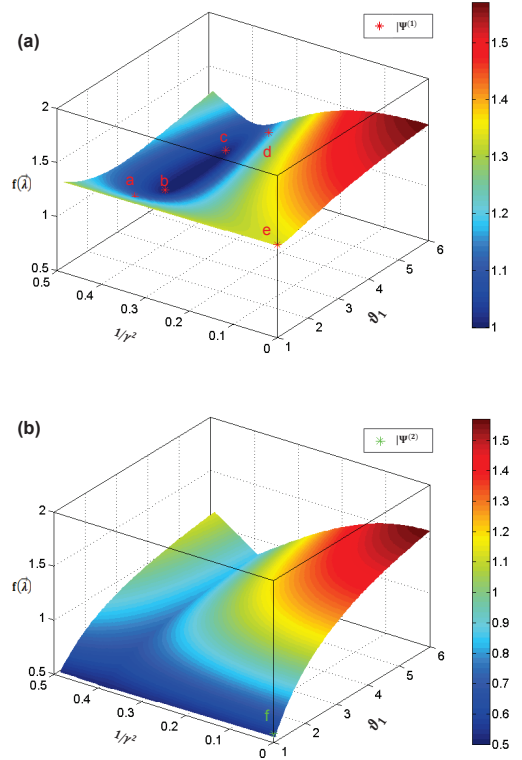}
\end{center}
\caption{Experimental and theoretical results for $f(\vec{\lambda})$
  as a function of $\theta_1$ and $1/\gamma^2$.  (a) shows the plot
  for the state $\ket{\Psi^{(1)}}$ together with the experimental data
  points `$a \thicksim e$'.  (b) shows the plot for the state
  $\ket{\Psi^{(2)}}$ and the data point '$f$' from the
  experiment. Error bars are too small to identify (see
  Table~\ref{tab:spectra}).}
\label{fig:result}
\end{figure}

The results are illustrated in Fig.~\ref{fig:resultp} and
\ref{fig:result}.  In Fig.~\ref{fig:resultp}, the data is shown in the
three-dimensional polytope for $\lambda_2^\text{max}$,
$\lambda_3^\text{max}$, $\lambda_4^\text{max}$ as by post-selection of
the first qubit, $\lambda_1^\text{max}=1$.  The smaller polytope
$\mathcal{P}^{s}$ becomes a three-dimensional polytope
$\tilde{\mathcal{P}}^{s}$ with vertices
$(1,1/2,1/2),(1/2,1,1/2),(1/2,1/2,1),(1,1,1)$, and the full polytope
$\mathcal{P}^{\text{full}}$ becomes a three-dimensional polytope
$\tilde{\mathcal{P}}^{\text{full}}$ which contains
$\tilde{\mathcal{P}}^{s}$ and has an additional vertex
$(1/2,1/2,1/2)$. The data point $f$ of the state $\ket{\Psi^{(2)}}$
outside of $\tilde{\mathcal{P}}^{s}$ shows that $\ket{\Psi^{(2)}}$ is
not in $\mathcal{P}^{s}$. In contrast, the data points $a,b,c,d,e$
obtained from $\ket{\Psi^{(1)}}$ all lie in $\tilde{\mathcal{P}}^{s}$,
which indicates that $\ket{\Psi^{(1)}}$ belongs to
$\mathcal{P}^{s}$. This shows that $\ket{\Psi^{(1)}}$ and
$\ket{\Psi^{(2)}}$ have different entanglement types.

In Fig \ref{fig:result}, the plot of $f(\vec{\lambda})$ as a function
of $\theta_1$ and $1/\gamma^2$ is shown. For the smaller polytope
$\mathcal{P}^{s}$, the inequality $f(\vec{\lambda})\geq 1$ always
holds. A violation of this inequality signals that the state
$\ket{\Psi^{(2)}}$ (point $f$) is not in $\mathcal{P}^{s}$. In
contrast, the data points $a,b,c,d,e$ obtained for $\ket{\Psi^{(1)}}$
all satisfy $f(\vec{\lambda})\geq 1$, which indicates that
$\ket{\Psi^{(1)}}$ belongs to $\mathcal{P}^{s}$. This shows that
$\ket{\Psi^{(1)}}$ and $\ket{\Psi^{(2)}}$ have different entanglement
types.

Since the first photon of the four-qubit state is post-selected and
the last photon goes through a non-unitary filter, the probability of
succes for the experiment is $0.2917$, $0.2222$, $0.1667$, $0.1768$,
$0.5$, and $0.5$ for our experimental data `$a \thicksim f$,'
respectively.

Because the birefringence of the $o$-light and the $e$-light in the
BBO (down-converter) cannot be compensated completely, and because of
the high-term noise from the SPDC process and some mode mismatch, we
do not obtain the pure states $\ket{\Psi^{(1)}}$ and
$\ket{\Psi^{(2)}}$, but some noisy version of them.  Nonetheless, the
errors in our experiment are mainly due to the time uncertainty of the
photon pairs generated in the BBO.  The coincidence counts obey a
Poisson distribution, the parameters of which we estimate from the
experimental data.  Then we perform Monte Carlo simulation to estimate
the errors indicated in Tables \ref{tab:4qubit} and \ref{tab:spectra}.

\textit{Summary --} We experimentally demonstrate the detection of
entanglement polytopes in a four-qubit system.  We use local filters
to effectively distinguish states with the same single-particle
spectra, but which belong to different polytopes.  This provides a new
tool to experimental detection of entanglement in a multi-qubit system
using only local operations.

\textit{Acknowledgements --} The work in USTC is supported by National
Fundamental Research Program (Grants No. 2011CBA00200 and
No. 2011CB9211200), National Natural Science Foundation of China
(Grants No. 61108009 and No. 61222504). B.Z. is supported by NSERC and
CIFAR.

\bibliography{Poly}

\newpage
\onecolumngrid

\section{Appendix}

\noindent{\bf A. Entanglement polytopes}

\noindent
Two quantum states $\ket{\psi_1}$ and $\ket{\psi_2}$ are said to be
equivalent with respect to SLOCC if there exists a sequence of local
operations and classical communication that converts the state
$\ket{\psi_1}$ into $\ket{\psi_2}$ with non-zero probability
$p_{1\to 2}>0$, and another protocol for the conversion of $\ket{\psi_2}$
into $\ket{\psi_1}$ that succeeds with probability $p_{2\to 1}>0$.
As we only require the success probabilities to be non-zero, it is
sufficient to consider one branch of the protocol that has non-zero
success probability. Thus we can, for example, write
\begin{equation}\label{eq:SLOCC2}
\sqrt{p_{1\to 2}}\ket{\psi_2}=M_1\otimes M_2\otimes\dots\otimes
  M_n\ket{\psi_1},
\end{equation}
where the matrices $M_i$ correspond to the combination of all
operations performed on particle $i$.  In \cite{DVC00} it was shown
that the matrices $M_i$ in (\ref{eq:SLOCC2}) can be replaced by
invertible matrices.  Moreover, all matrices can be chosen to have
determinant one, combining all scalar factors with the success
probability. Hence we have that the states $\ket{\psi_1}$ and
$\ket{\psi_2}$ are in the same SLOCC class if and only if there is a
non-zero constant $\lambda\in\C$ and matrices $A_i\in{\rm SL}(d_i)$ (where $d_i$ denotes the dimension of subsystem $i$) such that
\begin{equation}\label{eq:SL}
\lambda\ket{\psi_2}=A_1\otimes A_2\otimes\dots\otimes
  A_n\ket{\psi_1}.
\end{equation}
Results on the classification of pure four-qubit states with respect
to SLOCC can be found in \cite{VDDV02,CD06,LLSS07,LLHL09}.  Note that
in the literature, sometimes the scaling factor $\lambda$ is
incorrectly ignored.

Unlikely the situation for local unitary transformations, polynomial
invariants of the group ${\rm SL}(d)^{\otimes n}$ yield only a
necessary condition for SLOCC equivalence of two quantum states.
\begin{proposition}
Let $f_1,\ldots,f_m$ be homogeneous polynomial invariants of the group
${\rm SL}(d)^{\otimes n}$. If the normalized states $\ket{\psi_1}$ and
$\ket{\psi_2}$ are in the same SLOCC class, then there exists a
non-zero constant $\lambda\in\C$ such that
\begin{equation}\label{eq:SLOCC_condition}
f_i(\ket{\psi_1})=f_i(\lambda\ket{\psi_2})=\lambda^{\deg f_i}(\ket{\psi_2})
  \qquad\text{for $i=1,\ldots,m$}.
\end{equation}
\end{proposition}
In the case of four qubits, we have four polynomial invariants
$B_{0000}$, $D_{0000}^{(1)}$, $D_{0000}^{(2)}$, and $F_{0000}$ of
degree $2$, $4$, $4$, and $6$, respectively, see \cite{LuTh03}.

In order to get a necessary and sufficient criterion to decide SLOCC
equivalence, one may consider covariants.  Two vectors are in the same
orbit of the group ${\rm SL}(d)^{\otimes n}$ if and only if all
covariants agree.  Again, one has to take care of the scaling
parameter $\lambda$ to apply this criterion.  In the case of four
qubits, there are $170$ covariants, see \cite{BLT03}.

It has been shown that the points corresponding to the (sorted)
spectra of the single-particle reduced density matrices of pure
quantum states in the closure of an orbit under SLOCC transformations
form a so-called \emph{entanglement polytope}, see
\cite{WDGC13,SOK14}.  The vertices of the polytope correspond to the
covariants that do not vanish.

\bigskip

\noindent
{\bf B. Four-qubit Polytopes}

\begin{figure}[hbt!]
\centerline{\epsfxsize0.8\hsize\epsffile{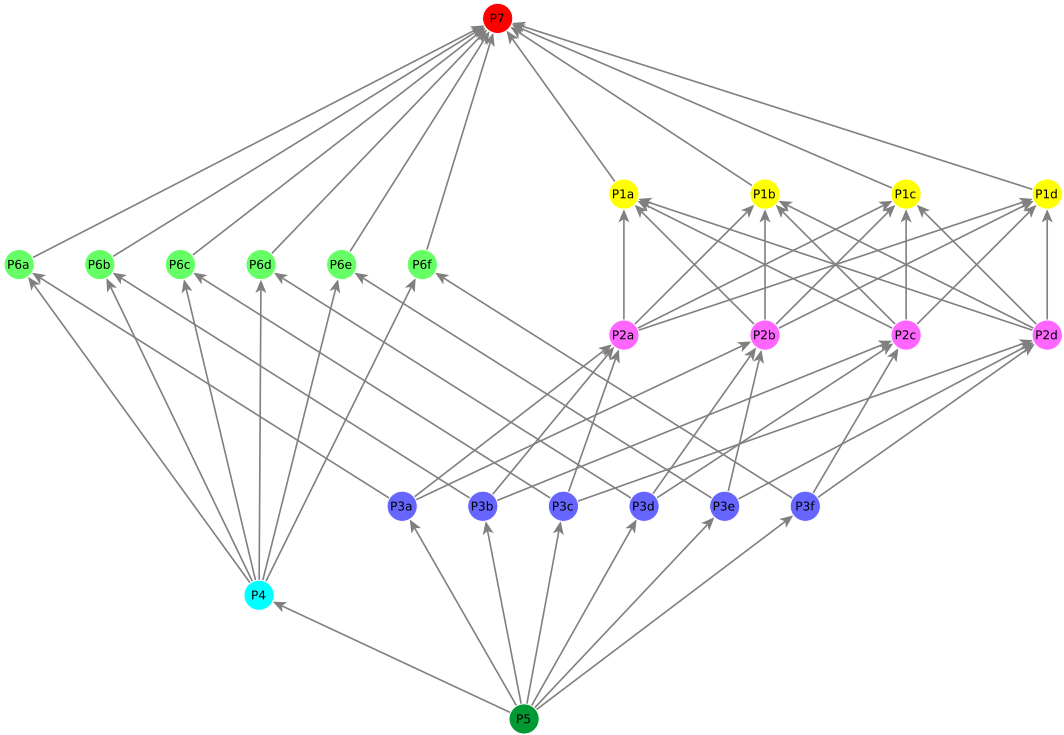}}
\caption{Lattice of the different entanglement polytopes for four
  qubits.  Note that, in general, the polytopes intersect
  non-trivially.}
\label{fig:polytope_lattice}
\end{figure}

\noindent
In the case of four qubits, there are $7$ different $4$-dimensional
polytopes up to permutation of the qubits \cite{WDGC13}.
Lower-dimensional polytopes correspond to states that partially
factorize.  When we also consider the permutations, the polytopes
$\mathcal{P}_3$ and $\mathcal{P}_6$ come in $6$ different versions,
while the polytopes $\mathcal{P}_1$ and $\mathcal{P}_2$ split into $4$
different subtypes. The vertices of all polytopes are listed in 
Table \ref{tab:polytopes}.
The polytope
$\mathcal{P}_5$ is contained in all other polytopes, and the polytope
$\mathcal{P}_7=\mathcal{P}^{\text{full}}$ is the largest polytope
containing all states. Fig. \ref{fig:polytope_lattice} illustrates how
the polytopes are contained in each other.

In the experiment, we investigate the four-qubit state
\begin{alignat}{5}
\ket{\Psi^{(1)}}={}&\frac{\sqrt{3}}{3}(\ket{HHHH}+\ket{VVVV})
 &{}+\frac{\sqrt{3}}{6}(\ket{HV}+\ket{VH})(\ket{HV}+\ket{VH}),
\end{alignat}
and the four-qubit GHZ state
\begin{equation}
\ket{\Psi^{(2)}}=\frac{\sqrt{2}}{2}(\ket{HHHH}+\ket{VVVV}).
\end{equation}
The qubits are encoded by horizontal $\ket{H}$ and vertical $\ket{V}$
polarization.  Evaluating the covariants from \cite{BLT03} for the
states $\ket{\Psi^{(1)}}$ and $\ket{\Psi^{(2)}}$ we find that the
corresponding polytopes are $\mathcal{P}^s=\mathcal{P}_4$ and the full
polytope $\mathcal{P}^{\text{full}}=\mathcal{P}_7$, respectively.  The
polytope $\mathcal{P}_4$ is obtained from $\mathcal{P}_7$ by removing
the vertex $(1/2,1/2,1/2,1)$ and all its permutations. The
discriminating inequalities are
\begin{alignat}{5}\label{eq:char_inequal}
f(\vec{\lambda})=-\lambda_1^{\text{max}}+\lambda_2^{\text{max}}+\lambda_3^{\text{max}}+\lambda_4^{\text{max}}\ge 1
\end{alignat}
and all its permutations.

\begin{table}[hbt!]
\footnotesize
\begin{tabular}{@{}|c|p{0.85\hsize}|@{}}
\hline
$\mathcal{P}_1^a$ &\rightskip0mm plus 50mm
$(  1,  1,  1,  1)$,
$(1/2,1/2,  1,  1)$,
$(1/2,  1,1/2,  1)$,
$(1/2,  1,  1,1/2)$,
$(  1,1/2,1/2,  1)$,
$(  1,1/2,  1,1/2)$,
$(  1,  1,1/2,1/2)$,
$(1/2,1/2,1/2,  1)$,
$(1/2,1/2,  1,1/2)$,
$(1/2,  1,1/2,1/2)$,
$(  1,1/2,1/2,1/2)$,
$(1/2,1/2,1/2,3/4)$\\
\hline
$\mathcal{P}_1^b$ &\rightskip0mm plus 50mm
$(  1,  1,  1,  1)$,
$(1/2,1/2,  1,  1)$,
$(1/2,  1,1/2,  1)$,
$(1/2,  1,  1,1/2)$,
$(  1,1/2,1/2,  1)$,
$(  1,1/2,  1,1/2)$,
$(  1,  1,1/2,1/2)$,
$(1/2,1/2,1/2,  1)$,
$(1/2,1/2,  1,1/2)$,
$(1/2,  1,1/2,1/2)$,
$(  1,1/2,1/2,1/2)$,
$(1/2,1/2,3/4,1/2)$\\
\hline
$\mathcal{P}_1^c$ &\rightskip0mm plus 50mm
$(  1,  1,  1,  1)$,
$(1/2,1/2,  1,  1)$,
$(1/2,  1,1/2,  1)$,
$(1/2,  1,  1,1/2)$,
$(  1,1/2,1/2,  1)$,
$(  1,1/2,  1,1/2)$,
$(  1,  1,1/2,1/2)$,
$(1/2,1/2,1/2,  1)$,
$(1/2,1/2,  1,1/2)$,
$(1/2,  1,1/2,1/2)$,
$(  1,1/2,1/2,1/2)$,
$(1/2,3/4,1/2,1/2)$\\
\hline
$\mathcal{P}_1^d$ &\rightskip0mm plus 50mm
$(  1,  1,  1,  1)$,
$(1/2,1/2,  1,  1)$,
$(1/2,  1,1/2,  1)$,
$(1/2,  1,  1,1/2)$,
$(  1,1/2,1/2,  1)$,
$(  1,1/2,  1,1/2)$,
$(  1,  1,1/2,1/2)$,
$(1/2,1/2,1/2,  1)$,
$(1/2,1/2,  1,1/2)$,
$(1/2,  1,1/2,1/2)$,
$(  1,1/2,1/2,1/2)$,
$(3/4,1/2,1/2,1/2)$\\
\hline
$\mathcal{P}_2^a$ &\rightskip0mm plus 50mm
$(  1,  1,  1,  1)$,
$(1/2,1/2,  1,  1)$,
$(1/2,  1,1/2,  1)$,
$(1/2,  1,  1,1/2)$,
$(  1,1/2,1/2,  1)$,
$(  1,1/2,  1,1/2)$,
$(  1,  1,1/2,1/2)$,
$(1/2,1/2,  1,1/2)$,
$(1/2,  1,1/2,1/2)$,
$(  1,1/2,1/2,1/2)$\\
\hline
$\mathcal{P}_2^b$ &\rightskip0mm plus 50mm
$(  1,  1,  1,  1)$,
$(1/2,1/2,  1,  1)$,
$(1/2,  1,1/2,  1)$,
$(1/2,  1,  1,1/2)$,
$(  1,1/2,1/2,  1)$,
$(  1,1/2,  1,1/2)$,
$(  1,  1,1/2,1/2)$,
$(1/2,1/2,1/2,  1)$,
$(1/2,  1,1/2,1/2)$,
$(  1,1/2,1/2,1/2)$\\
\hline
$\mathcal{P}_2^c$ &\rightskip0mm plus 50mm
$(  1,  1,  1,  1)$,
$(1/2,1/2,  1,  1)$,
$(1/2,  1,1/2,  1)$,
$(1/2,  1,  1,1/2)$,
$(  1,1/2,1/2,  1)$,
$(  1,1/2,  1,1/2)$,
$(  1,  1,1/2,1/2)$,
$(1/2,1/2,1/2,  1)$,
$(1/2,1/2,  1,1/2)$,
$(  1,1/2,1/2,1/2)$\\
\hline
$\mathcal{P}_2^d$ &\rightskip0mm plus 50mm
$(  1,  1,  1,  1)$,
$(1/2,1/2,  1,  1)$,
$(1/2,  1,1/2,  1)$,
$(1/2,  1,  1,1/2)$,
$(  1,1/2,1/2,  1)$,
$(  1,1/2,  1,1/2)$,
$(  1,  1,1/2,1/2)$,
$(1/2,1/2,1/2,  1)$,
$(1/2,1/2,  1,1/2)$,
$(1/2,  1,1/2,1/2)$\\
\hline
$\mathcal{P}_3^a$ &\rightskip0mm plus 50mm
$(  1,  1,  1,  1)$,
$(1/2,1/2,  1,  1)$,
$(1/2,  1,1/2,  1)$,
$(1/2,  1,  1,1/2)$,
$(  1,1/2,1/2,  1)$,
$(  1,1/2,  1,1/2)$,
$(  1,  1,1/2,1/2)$,
$(1/2,  1,1/2,1/2)$,
$(  1,1/2,1/2,1/2)$\\
\hline
$\mathcal{P}_3^b$ &\rightskip0mm plus 50mm
$(  1,  1,  1,  1)$,
$(1/2,1/2,  1,  1)$,
$(1/2,  1,1/2,  1)$,
$(1/2,  1,  1,1/2)$,
$(  1,1/2,1/2,  1)$,
$(  1,1/2,  1,1/2)$,
$(  1,  1,1/2,1/2)$,
$(1/2,1/2,  1,1/2)$,
$(  1,1/2,1/2,1/2)$\\
\hline
$\mathcal{P}_3^c$ &\rightskip0mm plus 50mm
$(  1,  1,  1,  1)$,
$(1/2,1/2,  1,  1)$,
$(1/2,  1,1/2,  1)$,
$(1/2,  1,  1,1/2)$,
$(  1,1/2,1/2,  1)$,
$(  1,1/2,  1,1/2)$,
$(  1,  1,1/2,1/2)$,
$(1/2,1/2,  1,1/2)$,
$(1/2,  1,1/2,1/2)$\\
\hline
$\mathcal{P}_3^d$ &\rightskip0mm plus 50mm
$(  1,  1,  1,  1)$,
$(1/2,1/2,  1,  1)$,
$(1/2,  1,1/2,  1)$,
$(1/2,  1,  1,1/2)$,
$(  1,1/2,1/2,  1)$,
$(  1,1/2,  1,1/2)$,
$(  1,  1,1/2,1/2)$,
$(1/2,1/2,1/2,  1)$,
$(  1,1/2,1/2,1/2)$\\
\hline
$\mathcal{P}_3^e$ &\rightskip0mm plus 50mm
$(  1,  1,  1,  1)$,
$(1/2,1/2,  1,  1)$,
$(1/2,  1,1/2,  1)$,
$(1/2,  1,  1,1/2)$,
$(  1,1/2,1/2,  1)$,
$(  1,1/2,  1,1/2)$,
$(  1,  1,1/2,1/2)$,
$(1/2,1/2,1/2,  1)$,
$(1/2,  1,1/2,1/2)$\\
\hline
$\mathcal{P}_3^f$ &\rightskip0mm plus 50mm
$(  1,  1,  1,  1)$,
$(1/2,1/2,  1,  1)$,
$(1/2,  1,1/2,  1)$,
$(1/2,  1,  1,1/2)$,
$(  1,1/2,1/2,  1)$,
$(  1,1/2,  1,1/2)$,
$(  1,  1,1/2,1/2)$,
$(1/2,1/2,1/2,  1)$,
$(1/2,1/2,  1,1/2)$\\
\hline
%
$\mathcal{P}_4$ &\rightskip0mm plus 50mm
$(  1,  1,  1,  1)$,
$(1/2,1/2,  1,  1)$,
$(1/2,  1,1/2,  1)$,
$(1/2,  1,  1,1/2)$,
$(  1,1/2,1/2,  1)$,
$(  1,1/2,  1,1/2)$,
$(  1,  1,1/2,1/2)$,
$(1/2,1/2,1/2,1/2)$\\
\hline
$\mathcal{P}_5$ &\rightskip0mm plus 50mm
$(  1,  1,  1,  1)$,
$(1/2,1/2,  1,  1)$,
$(1/2,  1,1/2,  1)$,
$(1/2,  1,  1,1/2)$,
$(  1,1/2,1/2,  1)$,
$(  1,1/2,  1,1/2)$,
$(  1,  1,1/2,1/2)$\\
\hline
$\mathcal{P}_6^a$ &\rightskip0mm plus 50mm
$(  1,  1,  1,  1)$,
$(1/2,1/2,  1,  1)$,
$(1/2,  1,1/2,  1)$,
$(1/2,  1,  1,1/2)$,
$(  1,1/2,1/2,  1)$,
$(  1,1/2,  1,1/2)$,
$(  1,  1,1/2,1/2)$,
$(1/2,  1,1/2,1/2)$,
$(  1,1/2,1/2,1/2)$,
$(1/2,1/2,1/2,1/2)$\\
\hline
$\mathcal{P}_6^b$ &\rightskip0mm plus 50mm
$(  1,  1,  1,  1)$,
$(1/2,1/2,  1,  1)$,
$(1/2,  1,1/2,  1)$,
$(1/2,  1,  1,1/2)$,
$(  1,1/2,1/2,  1)$,
$(  1,1/2,  1,1/2)$,
$(  1,  1,1/2,1/2)$,
$(1/2,1/2,  1,1/2)$,
$(  1,1/2,1/2,1/2)$,
$(1/2,1/2,1/2,1/2)$\\
\hline
$\mathcal{P}_6^c$ &\rightskip0mm plus 50mm
$(  1,  1,  1,  1)$,
$(1/2,1/2,  1,  1)$,
$(1/2,  1,1/2,  1)$,
$(1/2,  1,  1,1/2)$,
$(  1,1/2,1/2,  1)$,
$(  1,1/2,  1,1/2)$,
$(  1,  1,1/2,1/2)$,
$(1/2,1/2,  1,1/2)$,
$(1/2,  1,1/2,1/2)$,
$(1/2,1/2,1/2,1/2)$\\
\hline
$\mathcal{P}_6^d$ &\rightskip0mm plus 50mm
$(  1,  1,  1,  1)$,
$(1/2,1/2,  1,  1)$,
$(1/2,  1,1/2,  1)$,
$(1/2,  1,  1,1/2)$,
$(  1,1/2,1/2,  1)$,
$(  1,1/2,  1,1/2)$,
$(  1,  1,1/2,1/2)$,
$(1/2,1/2,1/2,  1)$,
$(  1,1/2,1/2,1/2)$,
$(1/2,1/2,1/2,1/2)$\\
\hline
$\mathcal{P}_6^e$ &\rightskip0mm plus 50mm
$(  1,  1,  1,  1)$,
$(1/2,1/2,  1,  1)$,
$(1/2,  1,1/2,  1)$,
$(1/2,  1,  1,1/2)$,
$(  1,1/2,1/2,  1)$,
$(  1,1/2,  1,1/2)$,
$(  1,  1,1/2,1/2)$,
$(1/2,1/2,1/2,  1)$,
$(1/2,  1,1/2,1/2)$,
$(1/2,1/2,1/2,1/2)$\\
\hline
$\mathcal{P}_6^f$ &\rightskip0mm plus 50mm
$(  1,  1,  1,  1)$,
$(1/2,1/2,  1,  1)$,
$(1/2,  1,1/2,  1)$,
$(1/2,  1,  1,1/2)$,
$(  1,1/2,1/2,  1)$,
$(  1,1/2,  1,1/2)$,
$(  1,  1,1/2,1/2)$,
$(1/2,1/2,1/2,  1)$,
$(1/2,1/2,  1,1/2)$,
$(1/2,1/2,1/2,1/2)$\\
\hline
$\mathcal{P}_7$ &\rightskip0mm plus 50mm
$(  1,  1,  1,  1)$,
$(1/2,1/2,  1,  1)$,
$(1/2,  1,1/2,  1)$,
$(1/2,  1,  1,1/2)$,
$(  1,1/2,1/2,  1)$,
$(  1,1/2,  1,1/2)$,
$(  1,  1,1/2,1/2)$,
$(1/2,1/2,1/2,  1)$,
$(1/2,1/2,  1,1/2)$,
$(1/2,  1,1/2,1/2)$,
$(  1,1/2,1/2,1/2)$,
$(1/2,1/2,1/2,1/2)$\\
\hline
\end{tabular}
\caption{The different entanglement polytopes for four qubits given by
  their vertices}
\label{tab:polytopes}
\label{tab:polytopes2}
\end{table}

\bigskip

\noindent
{\bf C. Volume of the Polytopes}

\noindent
In the case of three qubits, we have only two three-dimensional
polytopes $\mathcal{P}^W\subset\mathcal{P}^{GHZ}$ corresponding to the
SLOCC class containing the $W$-state and the $GHZ$-state,
respectively.  Picking a pure three-qubit state with respect to the
Haar measure at random, the resulting distribution of the eigenvalues
of the local density matrices has been computed in \cite{CDKW14}.
From this one finds that the volume of the sub-polytope $\mathcal{P}^W$
is $203/216\approx 93.98 \%$.  Hence the probability for a random
three-qubit state to have a local spectra corresponding to a point
outside the polytope $\mathcal{P}^W$ is only $13/216\approx 6.02 \%$.

For four qubits, we computed the local spectra of $10^6$ random pure
states and determined which of the polytopes contains the vector of
local spectra.  The results are summarized in Table
\ref{table:volume_polytopes}.  While the polytope $\mathcal{P}_4$
corresponding to the state $\ket{\Psi^{(1)}}$ of our experiment is
fairly low in the hierarchy of polytopes (see
Fig. \ref{fig:polytope_lattice}), the local spectra of only $9522$ out
of one million random states violate the discriminating inequalities
(\ref{eq:char_inequal}). Hence, the chance for a random four-qubit
state to have a local spectrum that lies outside of $\mathcal{P}_4$ is
only about $0.95 \%$.  This clearly indicates that one has to apply
local filters in order to get information about the entanglement
polytopes.

Note that after measuring one of the qubits and post-selection of the
measurement outcome, we have a four-qubit state that factors into a
single qubit and a three-qubit state.  The polytope $\mathcal{P}_4$ is
mapped to the three-qubit polytope $\mathcal{P}^W$ which has a volume
of about $94 \%$.  Hence the local measurement increases the chance
for a random state to lie outside the smaller polytope from less than
one percent to about six percent.

\begin{table}[hbt]\footnotesize
\[
\begin{array}{|c||r|}
\hline
\mathcal{P}_1^{\vphantom{d}} \\  996\,761 \\
\hline
\mathcal{P}_2^{\vphantom{d}} \\  863\,481 \\
\hline
\mathcal{P}_3^{\vphantom{f}} \\  781\,562 \\
\hline
\mathcal{P}_4 \\  990\,478 \\
\hline
\mathcal{P}_5 \\  130\,165 \\
\hline
\mathcal{P}_6^{\vphantom{f}} \\ 1\,000\,000 \\
\hline
\mathcal{P}_7 \\ 1\,000\,000 \\
\hline
\end{array}
\begin{array}{|c|c|c|c|c|c|}
\hline
\multicolumn{3}{|c|}{\begin{array}{c@{\quad}|@{\quad}c}
      \mathcal{P}_1^a & \mathcal{P}_1^b\\
            990\,140 & 990\,137
  \end{array}}&
\multicolumn{3}{|c|}{\begin{array}{c@{\quad}|@{\quad}c}
        \mathcal{P}_1^c & \mathcal{P}_1^d\\
              990\,204 & 990\,262\end{array}}\\
\hline
\multicolumn{3}{|c|}{\begin{array}{c@{\quad}|@{\quad}c}
        \mathcal{P}_2^a & \mathcal{P}_2^b\\
              705\,172 & 704\,928
  \end{array}}&
\multicolumn{3}{|c|}{\begin{array}{c@{\quad}|@{\quad}c}
        \mathcal{P}_2^c & \mathcal{P}_2^d\\
              704\,932 & 704\,791\end{array}}\\
\hline
\mathcal{P}_3^a & \mathcal{P}_3^b & \mathcal{P}_3^c & \mathcal{P}_3^d & \mathcal{P}_3^e & \mathcal{P}_3^f\\
607\,121 & 607\,010 & 607\,176 & 606\,791 & 606\,925 & 607\,051\\
\hline
\multicolumn{6}{|c|}{\mathcal{P}_4}   \\
\multicolumn{6}{|c|}{990\,478}\\
\hline
\multicolumn{6}{|c|}{\mathcal{P}_5}   \\
\multicolumn{6}{|c|}{130\,165}\\
\hline
\mathcal{P}_6^a & \mathcal{P}_6^b & \mathcal{P}_6^c & \mathcal{P}_6^d & \mathcal{P}_6^e & \mathcal{P}_6^f\\
995\,287 & 995\,277 & 995\,320 & 995\,158 & 995\,201 & 995\,191\\
\hline
\multicolumn{6}{|c|}{\mathcal{P}_7}\\
\multicolumn{6}{|c|}{1\,000\,000}\\
\hline
\end{array}
\]
\caption{Distribution of 1\,000\,000 random pure states on the different
  entanglement polytopes.  In the first column we list the union of the
different permuted polytopes of the same type.}
\label{table:volume_polytopes}
\end{table}

\end{document}